\documentclass{PoS}

\title{A new SPS programme}

\ShortTitle{A new SPS programme}

\author{{Marek Gazdzicki for the NA49-future Collaboration} \\
        University of Frankfurt, Frankfurt, Germany and Swietokrzyska Academy,
        Kielce, Poland \\
        E-mail: \email{Marek.Gazdzicki@cern.ch}}

\author{ \vspace*{-0.7cm} \begin{center} The NA49-future Collaboration 
\end{center} \\ }
\author{\vspace*{-0.5cm} N.~Antoniou, P.~Christakoglou,
        F.~Diakonos, A.~D.~Panagiotou, A.~Petridis, M.~Vassiliou \\
        University of Athens, Athens, Greece  
}
\author{ F.~Cafagna, M.~G.~Catanesi, T.~Montaruli, E.~Radicioni \\
University of Bari and INFN, Bari, Italy
}
\author{ D.~R\"ohrich \\
University of Bergen, Bergen, Norway
}
\author{L.~Boldizsar, Z.~Fodor, A.~Laszlo, G.~Palla, I.~Szentpetery,
        G.~Vesztergombi \\
KFKI Research Institute for Particle and Nuclear Physics
}
\author{J.~Cleymans \\
Cape Town University, Cape Town, South Africa
}
\author{J.~Brzychczyk, N.~Katrynska, R.~Karabowicz, Z.~Majka R.~Planeta, P.~Staszel \\
Jagellionian University, Cracow, Poland
}
\author{  B.~Baatar, V.~I.~Kolesnikov, A.~I.~Malakhov, G.~L.~Melkumov,
A.~N.~Sissakian, A.~S.~Sorin      \\
Joint Institute for Nuclear Research, Dubna, Russia
}
\author{  W.~Rauch      \\
Fachhochschule Frankfurt, Frankfurt, Germany
}
\author{  M.~Gazdzicki, B.~Lungwitz, M.~Mitrovski, R.~Renfordt,
T.~Schuster, C.~Strabel, H.~Stroebele      \\
University of  Frankfurt, Frankfurt, Germany
}
\author{  A.~Blondel, A.~Bravar, M.~Di~Marco      \\
University of Geneva, Geneva, Switzerland
}
\author{   J.~Blumer, R.~Engel, A.~Haungs, C.~Meurer, M.~Roth     \\
Forschungszentrum Karlsruhe, Karlsruhe, Germany
}
\author{  M.~Gazdzicki, R.~Korus, St.~Mr\'owczy\'nski, M.~Rybczynski,
P.~Seyboth, G.~Stefanek, Z.~Wlodarczyk,
A.~Wojtaszek      \\
\'Swi{\,e}tokrzyska Academy, Kielce, Poland
}
\author{    F.~Guber, A.~Kurepin, A.~Ivashkin, A.~Maevskaya    \\
Institute for Nuclear Research, Moscow, Russia
}
\author{   B.~Andrieu, J.~Dumarchez     \\
LPNHE, University of Paris VI and VII, Paris, France
}
\author{   K.-U.~Choi, J.-H.~Kim, J.-G.~Yi, I.-K.~Yoo     \\
an National University, Pusan, Republic of Korea
}
\author{   D.~Kolev,~R.~Tsenov     \\
Faculty of Physics, University of Sofia, Sofia, Bulgaria
}
\author{    A.~G.~Asryan, D.~A.~Derkach, G.~A.~Feofilov, S.~Igolkin, A.~S.~Ivanov,
R.~S.~Kolevatov, V.~P.~Kondratiev, P.~A.~Naumenko, V.~V.~Vechernin    \\
St. Petersburg State University, St. Petersburg, Russia
}
\author{   P.~Chung, R.~Lacey,  A.~Taranenko     \\
State University of New York, Stony Brook, USA
}
\author{  T.~Kobayashi, T.~ Nakadaira, K.~Sakashita, T.~ Sekiguchi      \\
Institute for Particle and Nuclear Studies, KEK,
Tsukuba,  Japan
}
\author{ K.~Grebieszkow, D.~Kikola, W.~Peryt, J.~Pluta, M.~Slodkowski, M.~Szuba  \\
Warsaw University of Technology, Warsaw, Poland
}
\author{  T. Anticic, K. Kadija, V. Nikolic, T. Susa      \\
Rudjer Boskovic Institute, Zagreb, Croatia
}

\abstract{ \vspace{-0.0cm}
A new experiemntal program to study hadron production in 
hadron-nucleus and nucleus-nucleus collisions at the CERN SPS
has been recently proposed by the NA49-future collaboration.
The physics goals of the program are:\\
-search for the critical point of strongly interacting matter
and a study of the properties of the onset of deconfinemnt
in nucleus-nucleus collisions,\\
-measurements of correlations, fluctuations and hadron spectra at 
 high $p_T$ in proton-nucleus collisions needed as for better
 understanding of nucleus-nucleus results,\\
-measurements of hadron production in hadron-nucleus interactions
needed for neutrino (T2K) and cosmic-ray (Pierre Auger Observatory and
KASCADE) expriments.
The physics of the nucleus-nucleus program is reviewed in this 
presentation. 
          }

\FullConference{Critical Point and Onset of Deconfinement\\
		 July 3-6 2006\\
		 Florence, Italy}

\begin{document}

\section{Introduction}

Recently The NA49-future
Collaboration has proposed \cite{proposal} to study hadron production in 
hadron-proton interactions and nucleus-nucleus 
collisions at the CERN SPS. 
This proposal follows the
Expression of Interest \cite{eoi} and the Letter of Intent 
\cite{loi} submitted
to the CERN SPS committee in November 2003 and January 2006, respectively.

The proposed physics program
consists of three subjects:\\

\vspace*{0.05cm}
\noindent
{\bf  - measurements of  hadron production in nucleus-nucleus collisions,
in particular fluctuations and long range correlations, with the
aim to identify the properties of the onset of deconfinement and
find evidence for the critical point of strongly interacting matter,}\\

\vspace*{0.05cm}
\noindent
{\bf  - measurements of hadron production in proton-proton and
proton-nucleus interactions 
needed as reference data for better understanding
of nucleus-nucleus reactions; in particular correlations, fluctuations
and high transverse momenta will be the focus of this study,  
}\\

\vspace*{0.05cm}
\noindent
{\bf - measurements of hadron production in hadron-nucleus
interactions needed for neutrino (T2K) and
cosmic-ray experiments (Pierre Auger Observatory and KASCADE).}\\

\noindent
It is foreseen to take data with proton and pion 
beams starting from 2007,
and with the beams of nuclei (C, Si and In) starting from 2009. 
The data taking period should end in 2011.
The invisaged run schedule is based on the assumption that the proposal
is approved not later than by the end of May 2007.

\noindent
The nucleus-nucleus program has the potential for an important
discovery -- the experimental observation of the critical point
of strongly interacting matter. We intend to carry out for the first
time in the history of heavy ion collisions a comprehensive
scan in two dimensional parameter space: size of colliding nuclei versus
interaction energy.
Other proposed studies belong to the class of precision
measurements. 

\noindent
The collaboration  proposes to perform these
measurements  
by use of the upgraded NA49 apparatus \cite{na49_nim}.
The most essential upgrades are
an increase of data taking and event rate by a factor of 24 and  
the construction of a projectile spectator detector
which will improve the accuracy of determination of the
number of projectile spectators by a factor of about 20.
The cost of all upgrades and detector maintenance is
estimated to be 1.5 MSFR.
Synergy of different physics programs as well as the use of
the existing accelerator chain and detectors offer the
unique opportunity 
to reach the  ambitious physics goals in a very
efficient and cost effective way.

\vspace*{0.2cm}
The NA49 apparatus at the  CERN~SPS served, during the last 10 years,
as a very reliable, large acceptance hadron spectrometer and
delivered high precision experimental data over the full
range of SPS beams (from proton to lead) \cite{na49_beam,Alt:2005zq}  
and energies (from
20$A$ GeV to 200$A$ GeV) \cite{afanasiev:2002mx,Gazdzicki:2004ef}.
Among the most important results from this study is the 
observation  \cite{afanasiev:2002mx,Gazdzicki:2004ef}
of narrow structures in the energy dependence of hadron production
in central Pb+Pb collisions.
These structures are located at the low CERN SPS energies
(30$A$--80$A$ GeV) and they are consistent with the predictions
\cite{Gazdzicki:1998vd}
for the onset of the deconfinement phase transition. 
The questions raised by this observation serve as a strong motivation 
for new measurements with nuclear beams in the SPS energy range
at the CERN~SPS as proposed by us and also envisaged at BNL~RHIC \cite{rhic_low}.
The proposed SPS and RHIC programs are to a large extent complementary
mainly due to different collision kinematics: 
beams on a fixed target at SPS and 
colliding beams at RHIC.  

\vspace*{0.2cm}
A report from the Villars meeting on ''Fixed-Target
Physics at CERN beyond 2005'' \cite{villars}
recognizes that the ion beams at the CERN SPS
remain ideal tools to study the features of the
phase transition between confined and deconfined
states of strongly interacting matter. It notes   
that an ion program aimed at the identification 
of the critical point and the study of its
properties is likely to be of substantial significance.
The NA49-future goals to
measure hadron production in hadron-nucleus
interactions needed for neutrino  and
cosmic-ray experiments as well as a search for the
critical point in nucleus-nucleus collisions were
summarized in the Briefing Book for European Strategy
for Particle Physics \cite{cern_strategy}.
The documents supporting the NA49-future physics program  
\cite{proposal} were provided by Frank Wilczek and 
by the T2K, Pierre Auger Observatory and 
KASCADE experiments.

\vspace*{0.2cm}
For the proposal in preparation
the NA49 apparatus and two detector prototypes were
tested in a 5-day long test run in August 2006 \cite{test_run}.
The performance of the NA49 TPCs has not shown any sign of
degradation since the beginning of their operation (1994).
In addition, this test clearly demonstrated the ability of the new
collaboration to operate the NA49 facility and the feasibility
of the proposed concepts of the new detectors.

\vspace*{0.2cm}
This presentation is based on the proposal of the
NA49-future Collaboration \cite{proposal} and  is limited to
the review  of the nucleus-nucleus part of the
physics program.

\section{Physics of Nucleus-Nucleus Collisions in NA49-future 
 }

\subsection{Key questions}
One of the key issues of contemporary physics is the understanding
of strong interactions and in particular the study of the
properties of strongly interacting matter in equilibrium.
What are the phases of this matter
and how do the transitions between them look like are questions
which motivate
a broad experimental and theoretical effort.
The study of high energy nucleus-nucleus collisions
gives us a unique possibility to address these questions
in well-controlled laboratory experiments.

\begin{figure}[!htb]
\includegraphics[width=0.45\textwidth]{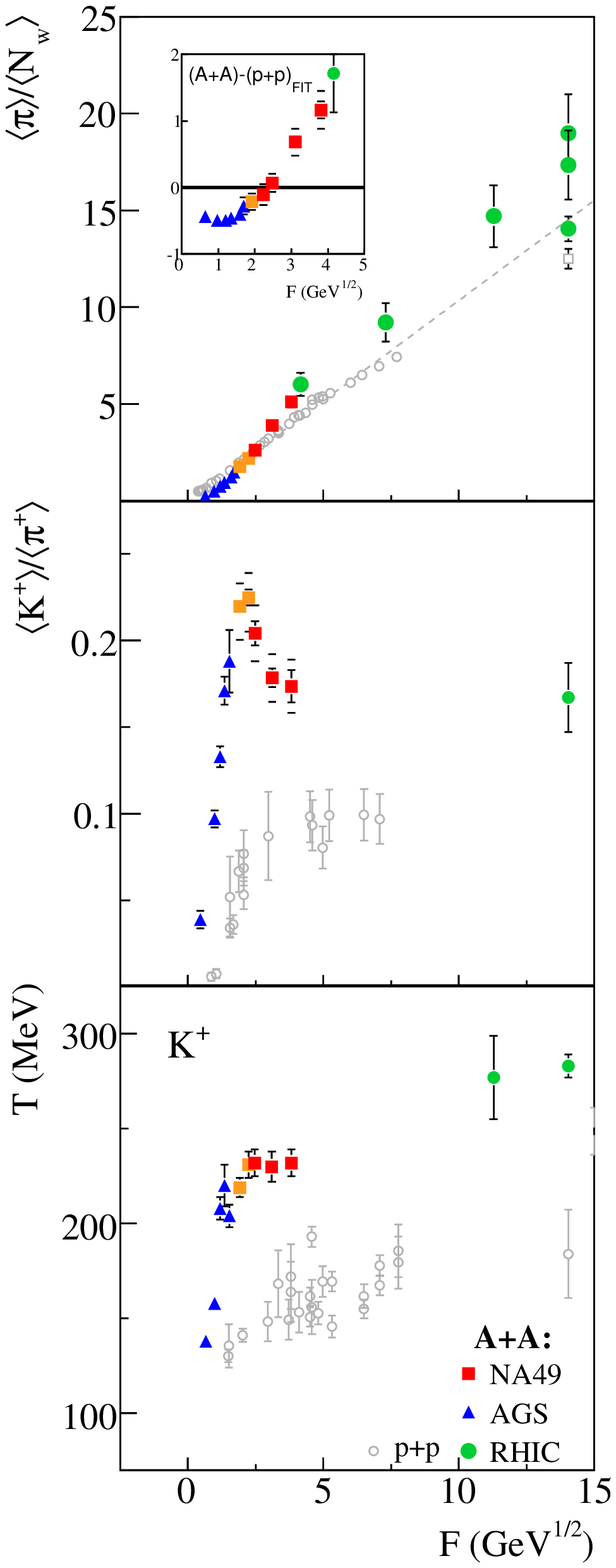}
\includegraphics[width=0.45\textwidth]{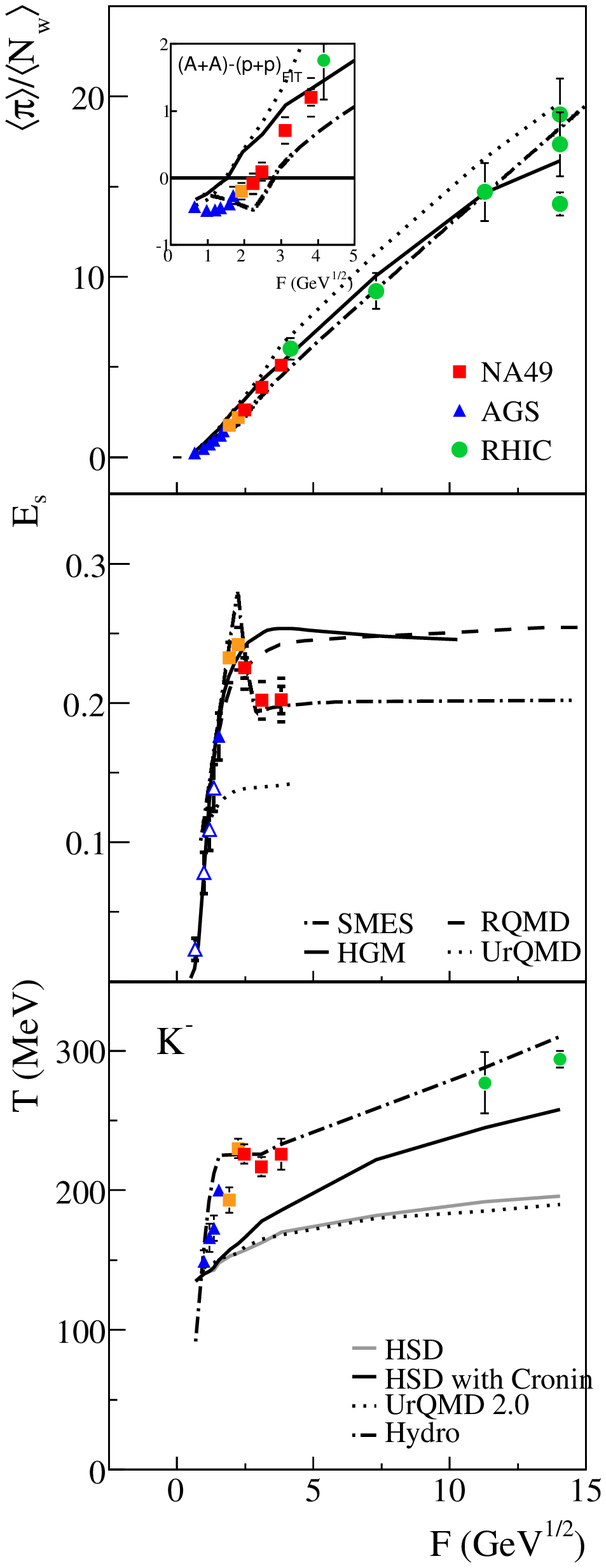}
\caption{\label{onset}
Left: Energy dependence of hadron production properties in central
Pb+Pb (Au+Au) collisions (closed symbols) and p+p interactions
(open symbols). Right: The results for Pb+Pb (Au+Au) reactions are
compared with various models. For details see text.
}
\end{figure}

\subsection{Onset of Deconfinement}

Recent results on the energy dependence of hadron production in central
Pb+Pb collisions at 20$A$, 30$A$, 40$A$, 80$A$ and 158$A$~GeV 
coming from the energy scan program at the
CERN SPS serve as evidence for the existence
of a transition
to a new form of strongly interacting matter, the Quark Gluon
Plasma (QGP) in nature.
Thus they are in agreement with the indications that at the top
SPS \cite{qgp_sps} and RHIC \cite{qgp_rhic}
energies the matter created at the early stage
of central Pb+Pb (Au+Au) collisions is in the state of QGP.

The key results and their comparison with models are summarized in
Fig.~\ref{onset}.
Energy dependence
of the mean pion multiplicity per wounded nucleon,
of the $\langle {\rm K}^+ \rangle/\langle \pi^+ \rangle$ ratio and
of the inverse slope parameter T of the transverse mass
spectra of K$^+$ mesons
measured in central Pb+Pb (Au+Au) collisions 
compared to results from p+p($\overline{\rm p}$) reactions 
are shown in Fig.~\ref{onset} (left).
The rapid changes
in the SPS energy range (solid squares) suggest the onset 
of  new physics in heavy ion collisions at the low SPS energies. 
Energy dependence of the mean pion multiplicity per wounded nucleon,
the relative strangeness production measured by the E$_S$ ratio
and the inverse slope parameter of
$m_T$--spectra of K$^-$ mesons
measured in central Pb+Pb (Au+Au) collisions
are compared
with predictions of various models in Fig.~\ref{onset} (right).
Models which do not assume the deconfinement phase transition
(HGM \cite{HGM}, RQMD \cite{RQMD}, UrQMD \cite{UrQMD} and HSD \cite{HSD}) 
fail to describe the data.
The introduction of a $1^{st}$ order phase transition at the
low SPS energies (SMES \cite{Gazdzicki:1998vd} 
and hydro \cite{Gorenstein:2003cu,brasil}) allows to describe the
measured structures in the energy dependence.

The energy dependence of the same observables measured in p+p interactions 
(open symbols in Fig.~\ref{onset} (left)) is
very different than that measured in central Pb+Pb (Au+Au)
collisions and  does not show any anomalies.

There are attempts to explain the results on nucleus-nucleus
collisions by (model dependent) extrapolations of the
results from proton-nucleus interactions \cite{fischer,cole}.
As detailed enough p+A data exist only at the top AGS and SPS
energies
the extrapolations are limited to these two energies and thus there are
no predictions concerning energy dependence of the quantities
relevant for the onset of deconfinement (see Fig.~\ref{onset}).
The underlying models are based on the assumption that particle yield in
the projectile hemisphere is due to production from
excited projectile nucleon(s).
This assumption is, however, in contradiction to the recent
results at RHIC \cite{bialas} and SPS \cite{mixing} energies
which clearly demonstrate a strong mixing of the projectile and
target nucleon contributions in the projectile hemisphere. 
Furthermore, qualitative statements on similarity or differences between
p+A and A+A reactions may be strongly misleading because of
a trivial kinematic reason. The center of mass system in 
p+A interactions moves toward the target nucleus A with increasing
A, whereas its position is A-independent for A+A collisions.
For illustration several typical examples are considered.
The baryon longitudinal momentum distribution in A+A collisions 
gets narrower with increasing A and, of course, it remains
symmetric in the collision center of mass system.
In p+A collisions it shrinks in the proton hemisphere and
broadens in the target hemisphere (for example see slide 5
in \cite{fischer}).
Thus a naive comparison would lead to the conclusion
that A+A collisions are qualitatively similar to p+A
interactions if the proton hemisphere is considered, or
that they are qualitatively different if the target hemisphere
is examined. 
One encounters similar difficulty in a discussion of
the A-dependence of particle ratios in a limited
acceptance.
For instance the kaon/pion ratio is independent of A
in p+A interactions if the total yields are considered
\cite{bialkowska}, it is however strongly A-dependent
in a limited acceptance (e.g. see slides 10 and 11  
in \cite{fischer}).
Recent string-hadronic models (\cite{RQMD, UrQMD, HSD})
take all trivial kinematic effects into account and
parametrize reasonably well p+p and p+A results.
Nevertheless they fail to reproduce the A+A data
(see e.g. Fig.~\ref{onset}).

\begin{figure}[!htb]
\begin{center}
\includegraphics[width=0.80\textwidth]{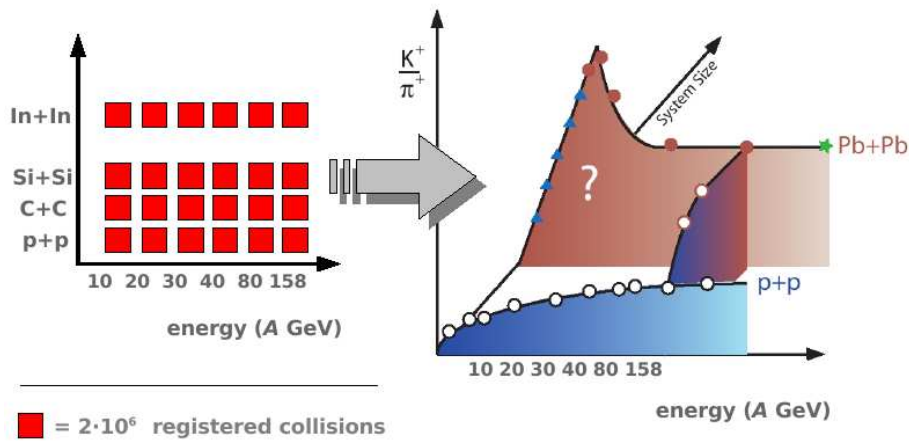}
\end{center}
\caption{\label{data3}
An illustration of the impact of the new measurements (of
central collisions) on clarifying the system size dependence of
the  $K^+/\pi^+$  anomaly observed in central Pb+Pb collisions
at the low SPS energies.
}
\end{figure}

Further progress in understanding effects which are
likely related to the onset of deconfinement can be done
only by a new comprehensive study of hadron production
in proton-nucleus and nucleus-nucleus collisions.

The two most important open questions are:
\begin{itemize}
\item
{\bf what is the nature of the transition from the anomalous energy dependence
measured in central Pb+Pb collisions at SPS energies to the smooth dependence
measured in p+p interactions?}
\item
{\bf is it possible to observe the predicted signals of the onset of
deconfinement in fluctuations \cite{mg_fluct}
and anisotropic flow \cite{Kolb:2000sd}?}
\end{itemize}

The qualitative progress in the experimental situation which will be
achieved by the proposed new measurements is illustrated in Fig.~\ref{data3}
using as an example the $K^+/\pi^+$ ratio. A detailed discussion
of the requested reactions is given below.

\subsection{Critical Point}

In the letter of Rajagopal, Shuryak, Stephanov and Wilczek
addressed to the SPS Committee one reads:
{\it ...  Recent theoretical developments suggest that a key
qualitative feature, namely a critical point
(of strongly interacting matter) which in a sense defines the
landscape to be mapped, may be within reach of discovery and analysis
by the SPS, if data is taken at several different energies.
The discovery of the critical point would in a stroke transform the map
of the QCD phase diagram which we sketch below from one based only on
reasonable inference from universality, lattice gauge theory and models
into one with a solid experimental basis. ...}
More detailed argumentation is presented below.

\begin{figure}[!ht]
\begin{center}
\includegraphics[width=0.80\textwidth]{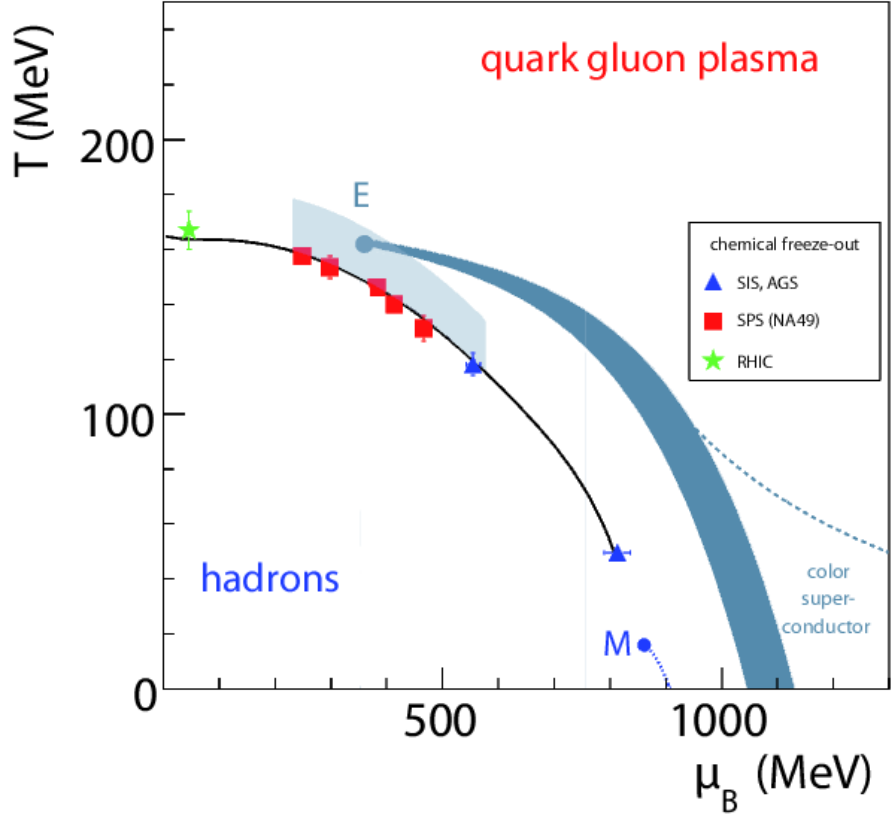}
\end{center}
\caption{The hypothetical phase diagram of strongly interacting
matter in the plane temperature, $T$, and baryonic chemical potential,
$\mu_B$. The end point {\bf E} of the first order transition strip is
the critical point of the second order.
The chemical freeze-out points extracted from the analysis
of hadron yields in central Pb+Pb (Au+Au) collisions at
different energies are plotted by the solid symbols.
The region covered by the future measurements at the CERN SPS
is indicated by the gray band.
}
\label{phase}
\end{figure}

Rich systematics of hadron multiplicities produced in nuclear collisions
can be described reasonably well by hadron gas models
\cite{Cleymans:1999cb,Braun-Munzinger:2003zd,becat}.
Among the model parameters fitted to the data are temperature, $T$, and baryonic 
chemical
potential, $\mu_B$, of the matter at the stage of freeze-out of the hadron
composition (the chemical freeze-out). These parameters extracted
for central Pb+Pb collisions at the CERN SPS energies are plotted in
the phase diagram of hadron matter,
Fig. \ref{phase}, together with the corresponding results for
higher (RHIC) and lower (AGS, SIS) energies. With increasing collision energy
the chemical freeze-out parameter $T$ increases and $\mu_B$ decreases.
A rapid increase of temperature is observed up to mid SPS energies, then from the
top SPS energy ($\sqrt{s_{NN}}$ = 17.2 GeV) to the top RHIC energy
($\sqrt{s_{NN}}$ = 200 GeV) the temperature increases only by about 10 MeV.

\begin{figure}[!ht]
\begin{center}
\includegraphics[width=0.80\textwidth]{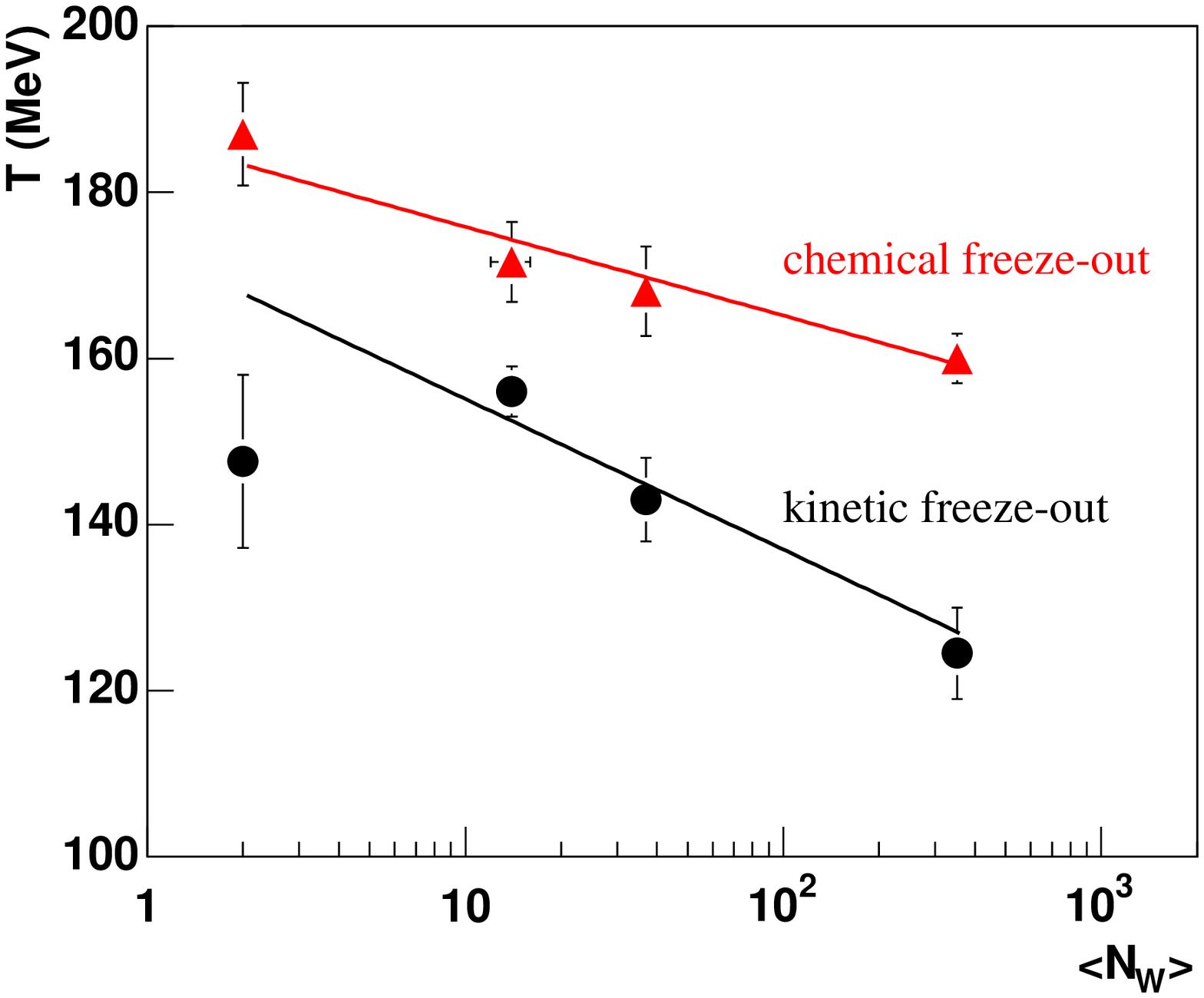}
\end{center}
\caption{The dependence of the chemical and kinetic freeze-out temperatures
on the mean number of wounded nucleons for p+p, C+C, Si+Si and
Pb+Pb collisions at 158$A$~GeV.
}
\label{temp_log}
\end{figure}

\vspace*{0.2cm}
Fig. \ref{phase} also shows a sketch of the phases of strongly interacting matter
in the ($T,\mu_B$) plane
as suggested by QCD-based considerations \cite{Rajagopal:2000wf, Stephanov:2004wx}.
To a large extent these predictions are qualitative, as
QCD phenomenology at finite temperature and baryon number
is one of the least explored domains of the theory.
More quantitative results come from
lattice QCD calculations which can be performed
at $\mu_B = 0$.
They suggest a rapid
crossover from the hadron gas to the QGP at the temperature
$T_C = 170-190$ MeV \cite{Karsch:2004wd,katz}, which seems to be somewhat higher 
than the
chemical freeze-out temperatures of
central Pb+Pb collisions ($T =150-170$ MeV) \cite{jakko}
at the top SPS and RHIC energies.

The nature of the transition to QGP is expected
to change with increasing baryo-chemical potential.
At high potential the transition may  be of the
first order, with the end point of the first order transition
domain, marked $E$ in Fig.~\ref{phase},
being the critical point of the second order.
Recently even richer structure of the phase transition to QGP
was discussed within a statistical model of quark-gluon bags
\cite{Gorenstein:2005rc}.
It was suggested that the line of the first order phase transition
at high $\mu_B$ is followed by a line of second order
phase transition at intermediate $\mu_B$, and then by lines of
''higher order transitions'' at low $\mu_B$.
A characteristic property of the second order phase transition
(the critical point or line) is a
divergence of the susceptibilities.
Consequently
an important signal of a second-order phase transition
at the critical point are large fluctuations, in particular
an enhancement of fluctuations of multiplicity and transverse
momentum are predicted \cite{Stephanov:1999zu}.
A characteristic feature of the second order phase transition 
is the validity
of appropriate power laws in measurable quantities related to critical fluctuations.
Techniques associated with such measurements in nuclear collisions
are under  development  \cite{antoniou:2005}
with emphasis on the sector of isoscalar di-pions ($\sigma$-mode)
as required by the QCD conjecture for the critical end point
in quark matter \cite{Rajagopal:2000wf}.
Employing such techniques in a study of nuclear collisions
at different energies at the SPS and with nuclei of different sizes,
the experiment may test not only the existence
and location of the critical point but also the size of critical
fluctuations as given by the critical exponents of the QCD conjecture.

Thus when scanning the phase diagram a maximum
of fluctuations located in a domain close to the critical point
(the increase of fluctuations can be expected over
a region $\Delta T \approx 15$ MeV and
$\Delta \mu_B \approx 50$ MeV \cite{Hatta:2002sj}) or the
critical line
should signal the second order phase transition.
The position of the critical region is uncertain,
but the best theoretical estimates based on lattice
QCD calculations locate it at $T \approx 158$ MeV and
$\mu_B \approx 360$ MeV
\cite{Fodor:2004nz,Allton:2005gk} as indicated in Fig. \ref{phase}.
It is thus in the vicinity  of the chemical freeze-out points
of central Pb+Pb collisions at the CERN SPS energies.

\begin{figure}[!ht]
\begin{center}
\includegraphics[width=0.60\textwidth]{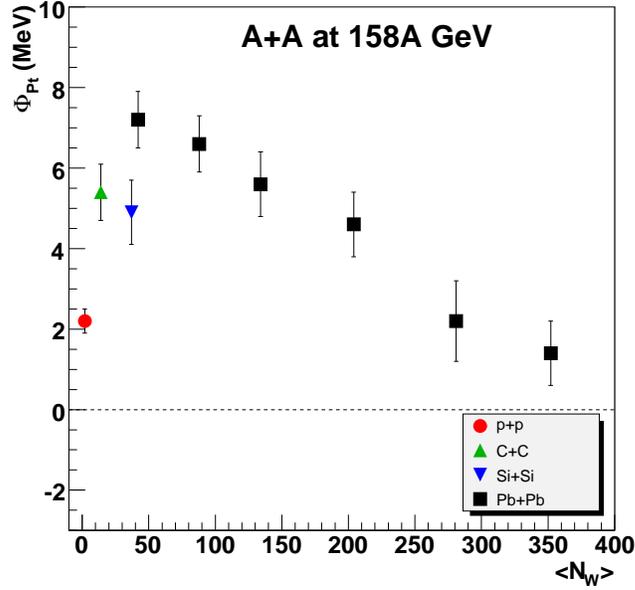}
\end{center}
\caption{The measure of transverse momentum fluctuations
$\Phi_{p_T}$ versus mean number of interacting nucleons
$\langle N_W \rangle$ for nuclear collisions at 158$A$ GeV.
The results for all charged
hadrons are presented.
Only statistical errors are plotted, the systematic
errors are smaller than 1.6 MeV.}
\label{phipt}
\end{figure}

Pilot data \cite{na49_beam} on interactions of light nuclei (Si+Si, C+C and p+p)
taken by NA49 at 40$A$ and 158$A$ GeV indicate that the freeze-out
temperature increases with decreasing mass number, $A$, of the
colliding nuclei, see Fig.~\ref{temp_log}.
This means that a scan in the collision energy and
mass of the colliding nuclei allows us to scan  the
($T,\mu_B$) plane in a search for the critical
point (line) of strongly interacting matter
\cite{Stephanov:1999zu}.

The experimental search for the critical point by investigating
nuclear collisions is justified at energies higher
than the energy of the onset of deconfinement.
This is because the energy density at the early stage of
the collision, which is required for the onset of deconfinement
is higher than the energy density at freeze-out, which
is relevant for the search for the critical point.
The only anomalies possibly related to the onset of deconfinement
are measured at 30$A$ GeV ($\sqrt{s_{NN}} \approx 8$ GeV)
(see Fig.~\ref{onset}). This limits
a search for the critical point to an energy range
$E_{lab} > 30A$ GeV ($\mu_B(CP) < \mu_B(30A \rm{GeV})$).

Fortunately, as discussed above and illustrated in Fig.~\ref{phase},
the best theoretical predictions locate the  critical point
in the ($T,\mu_B$)  region
accessible in nuclear collisions in this energy range 
(at about 80$A$ GeV). There are, however, large and difficult
to estimate systematic errors in these predictions.

Under the minimal assumption that the critical point is
located with equal probability at $\mu_B(CP) < \mu_B(30A \rm{GeV})$
one can estimate a probability of 0.5 that it is reachable in the
SPS energy range. 

One of the first proposed signals of the critical point of
strongly interacting matter was a maximum of the transverse
momentum fluctuations in the (collision energy)-(system size)
plane \cite{Stephanov:1999zu}.
NA49 performed the system size scan at 158$A$ GeV and,
in fact, a maximum was observed for collisions
with a number of interacting nucleons of about 40
\cite{na49_pt}.
Qualitatively similar results were obtained by CERES 
\cite{ceres}.
The experimental results of NA49 are shown in Fig.~\ref{phipt}, where
the intensive fluctuation measure, $\Phi_{P_T}$ is plotted
against  the mean number of interacting nucleons.
The magnitude of the observed fluctuations,
$\Phi_{P_T} = 7 \pm 2$ MeV, is in approximate agreement
with the predictions for the critical point,
$\Phi_{P_T} \approx 10$ MeV \cite{Stephanov:2001zj}.

\begin{figure}[!ht]
\begin{center}
\includegraphics[width=0.80\textwidth]{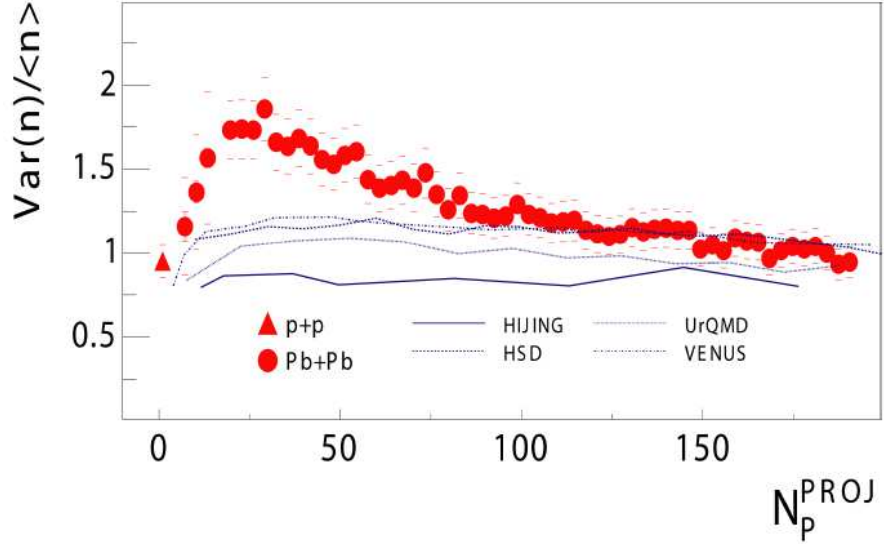}
\end{center}
\caption{
Scaled variance of the multiplicity distribution for
negatively charged hadrons as a function of the number
of projectile participants for Pb+Pb collisions at 158$A$ GeV.
The systematic errors due to a poor resolution in the
measurement of the number of projectile spectators
are indicated by the horizontal bars.
The lines show predictions of string-hadronic models.
}
\label{mult_fluct_models}
\end{figure}

Onset of deconfinement and the critical point are also expected
to lead to anomalies in the 
multiplicity fluctuations~\cite{mg_fluct,Stephanov:1999zu}.
The dispersion of the multiplicity distribution
should increase by about 10-20\% when crossing
the energy domain in which the onset of deconfinement is
located~\cite{mg_fluct}. 
The scaled variance of the multiplicity distribution is
expected to increase by about 10\% in the vicinity of the
critical point \cite{Stephanov:1999zu}. 
The identification of these  effects is however non-trivial.
This is mainly because 
the measured multiplicity fluctuations are directly sensitive
to the fluctuations in the number of interacting nucleons caused
by  event-by-event changes in the collision geometry. 

\begin{figure}[!ht]
\begin{center}
\includegraphics[width=0.80\textwidth]{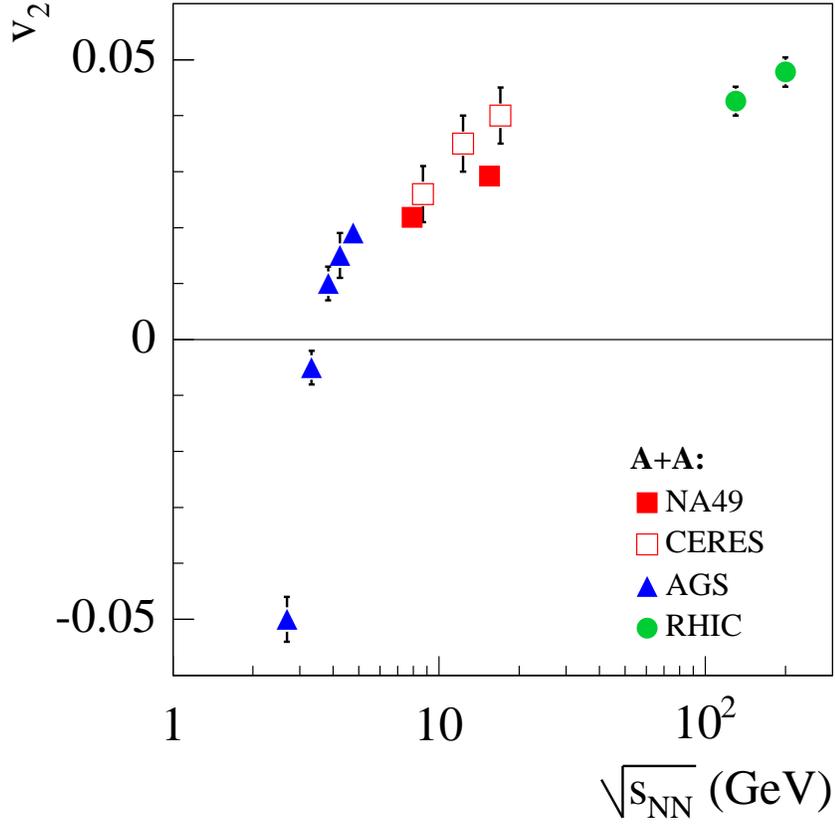}
\end{center}
\caption{
The energy dependence of
elliptic flow of charged pions at mid-rapidity
in Pb+Pb (Au+Au) collisions for mid-central events.
}
\label{v2edep}
\end{figure}

First results from NA49 on the scaled variance
of the multiplicity distribution as a function of the number of interacting nucleons from
the projectile nucleus are shown in Fig.~\ref{mult_fluct_models}.
The measurements show an increase towards peripheral collisions
similar to that seen for $\Phi_{P_T}$. 
This increase is not present in existing string-hadronic models
\cite{soft_balls_1}. 
It was recently suggested~\cite{soft_balls} that
the large fluctuations for peripheral collisions may  be caused
by equilibration (mixing) of particle sources from the projectile and target nuclei.
It is clear that clarification of the origin of the observed 
large increase of multiplicity fluctuations in peripheral collisions is 
a necessary first step in 
a search for the fluctuation signals of 
the critical point or onset of deconfinement. This requires new high precision 
comprehensive data on
multiplicity fluctuations and an improved forward calorimeter 
to reduce uncertainties
in the determination  of the collision geometry.

Whether the measured $p_T$ and multiplicity fluctuations signal the vicinity of
the critical point (line) remains an open question.
This is mainly because systematic data on the energy and collision system size dependence
of these fluctuation observables are missing. Only an  observation
of a maximum (or an onset) in the energy dependence will serve
as a strong indication for the critical point (line).

\begin{figure}[!ht]
\begin{center}
\includegraphics[width=0.80\textwidth]{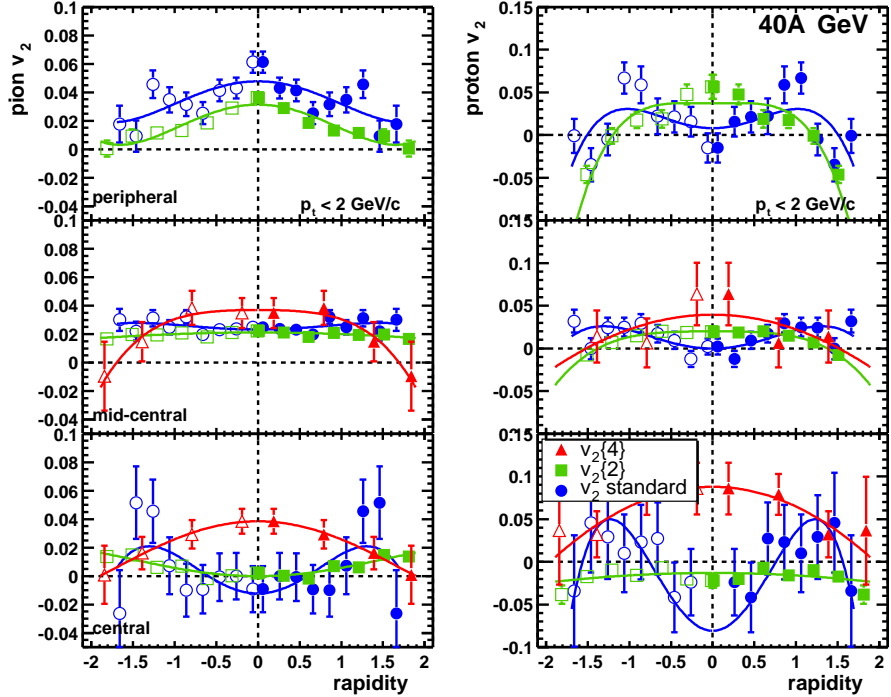}
\end{center}
\caption{
Elliptic flow of charged pions (left) and protons (right) obtained from
the standard ($v_2$ standard) and cumulant ($v_2$\{2\}) methods as a function of the
rapidity in Pb+Pb collisions at 40$A$ GeV for three centrality bins.
The open points have been reflected with respect to mid-rapidity.
The solid lines are from polynomial fits.
}
\label{na49-flow}
\end{figure}

The energy dependence of anisotropic flow is considered 
to be sensitive to both the onset of deconfinement \cite{Kolb:2000sd}
and the  critical point \cite{Shuryak:2005vk}. 
The two can be distinguished by separate measurements of
the flow for mesons and baryons.
In the case of the onset of deconfinement the flow of both
mesons and baryons should be reduced \cite{Bratkovskaya:2004nd}, 
whereas the critical point
should lead to a decrease of the baryon flow and an increase
of the meson flow \cite{Shuryak:2005vk}.
However, the existing
data \cite{na49-flow} are inconclusive on
whether the expected effects are present in the CERN SPS
energy range.
The main experimental results 
are summarized in Figs.~\ref{v2edep} and~\ref{na49-flow}.
The energy dependence of the $v_2$ parameter for pions in Pb+Pb (Au+Au)
collisions is shown in Fig.~\ref{v2edep}. A rapid increase observed at low
energies seems to weaken in the SPS energy range.  
In Fig.~\ref{na49-flow}
 the rapidity dependence of the elliptic flow parameter,
$v_2$, of pions and protons is plotted for Pb+Pb
collisions at 40$A$ GeV.
The results from the standard analysis suggest
the reduction of $v_2$ for protons  at mid-rapidity. 
This effect is, however, not observed in the results from the 
cumulant analysis and the 40$A$ GeV  data from NA49 
are the only data on proton flow at low SPS energies.
Thus the present data suggest the possibility of
anomalies in the energy dependence of elliptic flow of
mesons and baryons at the SPS energies, but they are too
sparse to allow any firm conclusion. 

It is thus clear that only the new measurements 
at the CERN SPS can answer the
important question:
\begin{itemize}
\item
{\bf does the critical point of strongly interacting matter
exist in nature and, if it does, where is it located?}
\end{itemize}

\begin{figure}[!htb]
\begin{center}
\includegraphics[width=0.80\textwidth]{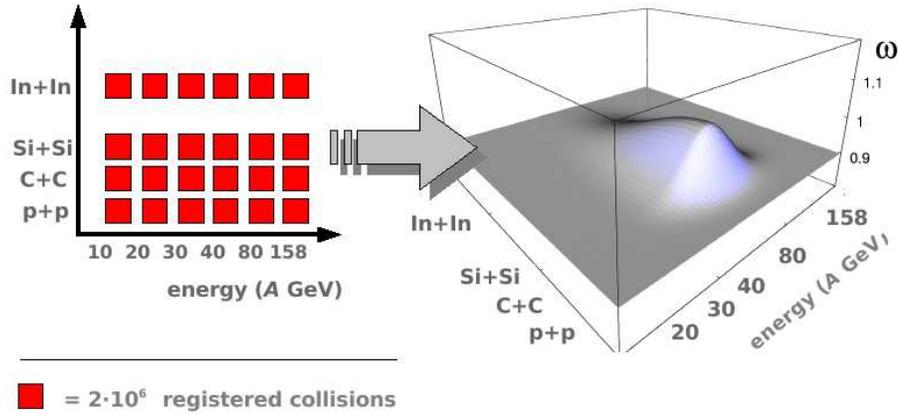}
\end{center}
\caption{\label{data2}
An illustration of the impact of the new measurements of central collisions
on the search for the critical point of strongly interacting matter.
}
\end{figure}

The qualitative progress in the experimental search which will be
achieved by the proposed new measurements is illustrated in Fig.~\ref{data2}.
The critical point is expected to manifest itself e.g. by a maximum of
the scaled variance $\omega$ of the produced
particle multiplicity distribution. A detailed discussion
of the requested reactions is given below.

\vspace*{0.2cm}
In conclusion, the recent experimental and theoretical findings
strongly suggest that a further study of nuclear collisions
in the CERN SPS energy range is of particular importance.
The new measurements will dramatically  improve the current
experimental situation and they should
allow to answer the key questions concerning the nature
of the onset of deconfinement and the existence and location
of the critical point.

\subsection{General requirements}

The physics goals of the new experimental program with
nuclear beams at the CERN SPS presented in the previous
section require a comprehensive energy scan in the whole SPS energy
range (10$A$-200$A$ GeV) with light and intermediate mass
nuclei. 
The NA49-future collaboration intends to register p+p,
C+C, Si+Si and In+In collisions at 10$A$,
20$A$, 30$A$, 40$A$, 80$A$, 158$A$ GeV and a typical
number of recorded events per reaction and energy of $6 \cdot 10^6$.

\begin{figure}[!htb]
\begin{center}
\includegraphics[width=0.80\textwidth]{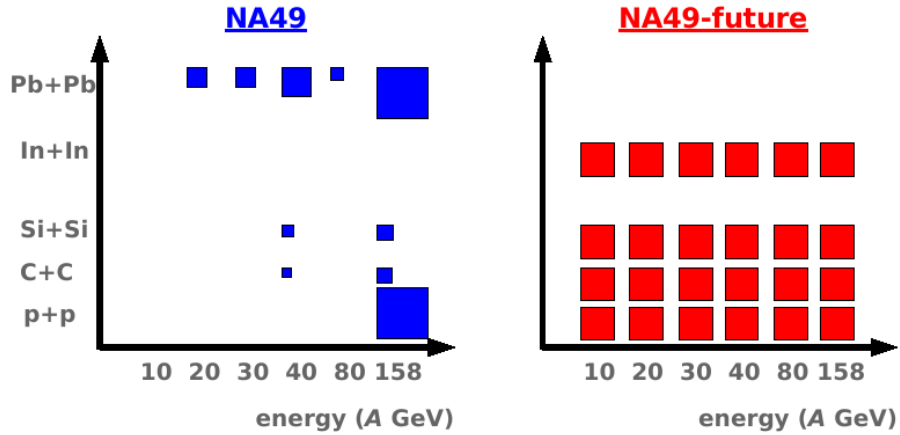}
\end{center}
\caption{\label{data1}
The  data sets recorded by NA49 and those planned to be
recorded by NA49-future.
The area of the boxes is proportional to the number of
registered central collisions, which for NA49-future will be
$2 \cdot 10^6$ per reaction.
}
\end{figure}

The  data sets recorded by NA49 and those planned to be
recorded by NA49-future are shown in Fig.~\ref{data1}. 
It is clear that the new
measurements will lead to a very significant experimental
progress.

\begin{figure}[!htb]
\begin{center}
\includegraphics[width=0.38\textwidth]{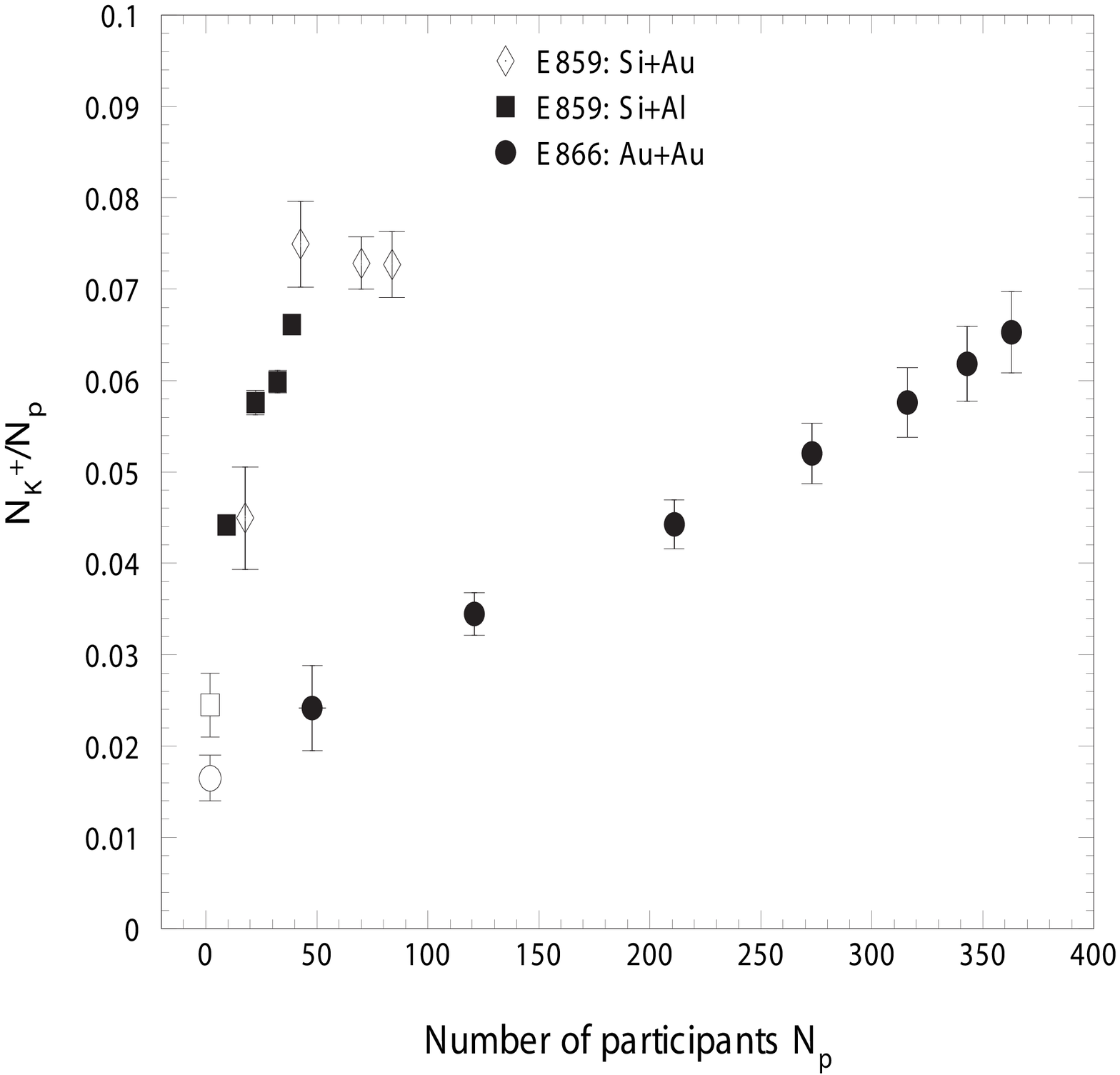}
\includegraphics[width=0.57\textwidth]{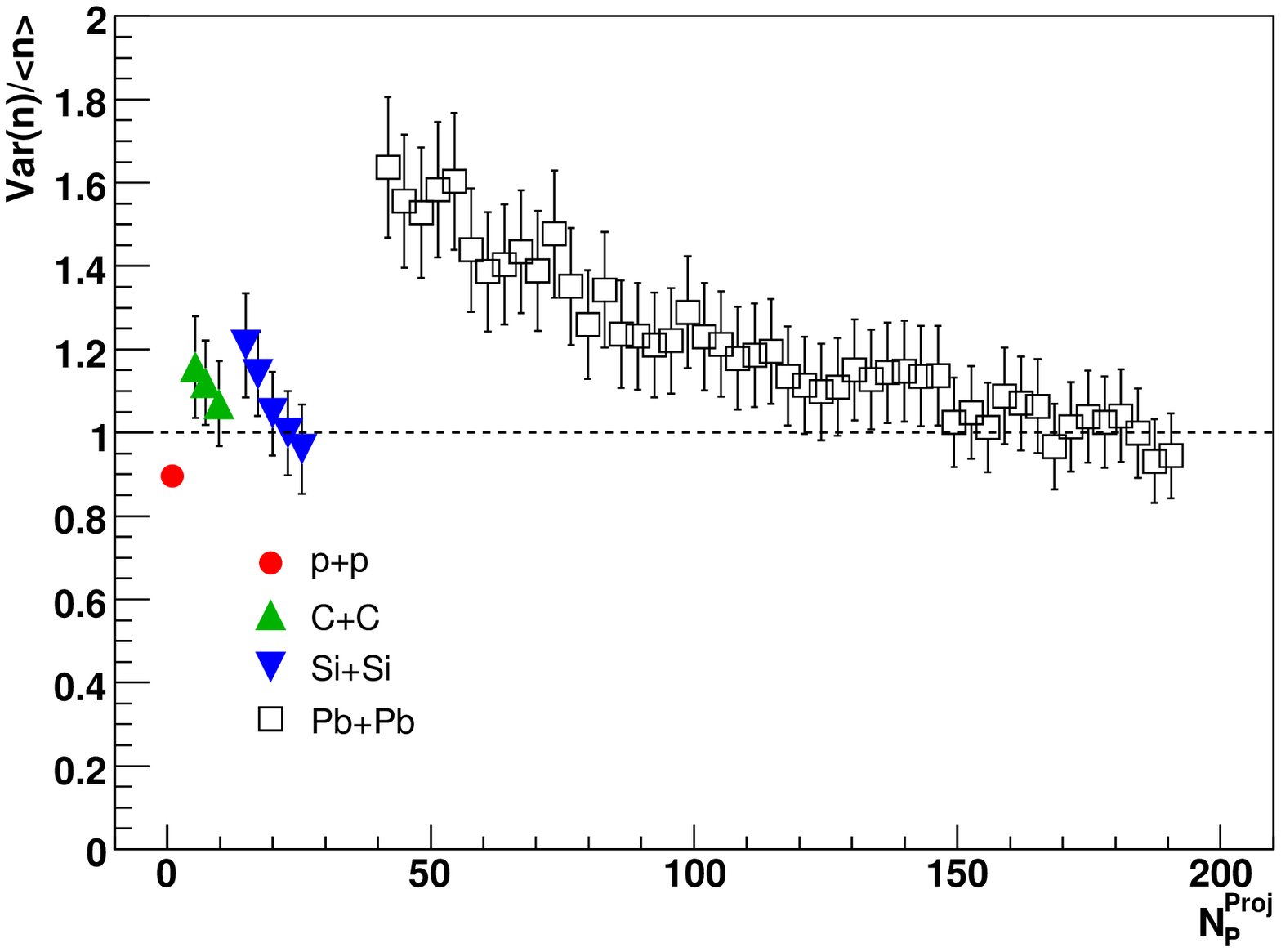}
\end{center}
\caption{\label{central} 
Left: The mean number of K$^+$ mesons per participant nucleon
is plotted as a function of the number of participants for
for Si+Al, Si+Au and Au+Au collisions at the top AGS energies
(10$A$-14$A$ GeV).
Right: The scaled variance of the multiplicity distribution 
of negatively charged hadrons in the
projectile hemi-sphere for p+p, C+C, Si+Si and Pb+Pb 
interactions is plotted versus the number of projectile 
participants. 
}
\end{figure}

It is important to underline that collisions of medium size
and light nuclei can not be replaced by centrality selected
collisions of heavy (e.g. Pb+Pb) nuclei. This point is
illustrated in Fig.~\ref{central}, where a large difference
between hadron production properties in central collisions
of light nuclei and peripheral Pb+Pb (Au+Au) interactions
is clearly seen. This conclusion is valid both for mean values
(Fig.~\ref{central} (left)) and for fluctuations 
(Fig.~\ref{central} (right)). 
Furthermore it has to be stressed that a hadronic final state produced
in central collisions is much closer to the chemical and thermal
equilibrium than a corresponding final state produced in peripheral
collisions. Thus, for the planned study of the properties of the
onset of deconfinement and the search for the critical point
central collisions are of primary interest.

The future analysis should  focus on the
study of fluctuations and anisotropic flow.
The first NA49 results on these subjects \cite{na49-flow,na49_pt,
na49_n,na49_k} suggest, in fact,
the presence of interesting effects for collisions with
moderate number of participant nucleons and/or low
collision energies.
However, as discussed above, a very limited set of data and  resolution 
limitations do not allow firm conclusions.

Several upgrades 
of the current NA49 apparatus are necessary to reach
the physics goals.
\begin{enumerate}
\item
The event collection rate should be significantly increased
in order to allow a fast collection of sufficient statistics
for a large number of different reactions ($A$, $\sqrt{s_{NN}}$).
\item
The resolution in the event centrality determination based
on the measurement of the energy of projectile spectator
nucleons should be improved.
This is important for  high precision measurements of
event-by-event fluctuations.
\item
It is desirable to increase
the acceptance for the measurements of charged hadrons.
\end{enumerate}

The  proposed hardware upgrades
of the NA49 apparatus 
and the resulting improvements
of its physics performance are presented in the proposal \cite{proposal}.

\subsection{Experimental landscape}

The SPS energy range is of particular importance
for the study of nucleus-nucleus collisions for
two reasons.
Firstly, at the low SPS energies anomalies in
the energy dependence of hadron production properties
are observed.
They are attributed to the onset of deconfinement.
Secondly, at high SPS energies the critical point of
strongly interacting matter can be discovered. This
expectation is based on the recent estimate of the
location of the critical point from lattice QCD
and the freeze-out parameters determined from 
measured hadron yields using the hadron-resonance gas model.

Among other existing heavy ion accelerators only RHIC at BNL
can potentially run in the SPS energy range.
In March 2006 a RIKEN BNL workshop 
''Can the QCD critical point be discovered at the BNL RHIC''
took place, where it was decided to start preparations
for a low energy RHIC program. 
The first tests of the accelerator complex  were
performed in June 2006. Proton beams of 11 GeV/c ($\sqrt{s_{NN}} = 22$~GeV)
were successfully injected, circulated and collided at RHIC.
The magnet currents for this test correspond to those required
for Au+Au running at $\sqrt{s_{NN}} \approx 8.8$~GeV,
or equivalently with fixed target running at about 40$A$ GeV.
Tests at lower energies using ion beams are foreseen in 2007.
The physics run with Au beams at several energies 
which correspond to the NA49 settings may be scheduled
for 2009. 
The performance of the STAR and PHENIX detectors matches,
in general, the physics requirements of running at low RHIC
energies.

The SPS program proposed here and the suggested new
RHIC program are to a large extent complementary.
This is due to different collision kinematics and different
priorities in data taking.

NA49-future at the SPS will take data in the fixed target mode.
First, this gives the unique possibility to measure the
number of projectile spectators for each collision, 
which, in turn, allows a precise study of event-by-event
fluctuations, in particular, fluctuations of extensive
quantities like multiplicity.
Second, the NA49-future acceptance for identified hadrons
covers a large part of the projectile hemi-sphere.
This allows measurement of almost the full rapidity
spectrum and total multiplicities of produced hadrons. 
NA49-future intends to perform the energy scan for collisions
of light and medium size nuclei (p+p, C+C, Si+Si and In+In).

In contrast, it is planned to start the low energy RHIC program with the
energy scan for Au+Au collisions.
STAR and PHENIX at the RHIC will take data in the collider mode.
This allows to perform measurements with full and 
uniform azimuthal angle acceptance and forward--backward
symmetry with respect to mid-rapidity.
This is crucial for a precise measurement of 
azimuthal flow and other measures of azimuthal
correlations.

Starting from 2015 the study of nucleus-nucleus collisions at 
lower SPS energies (10$A$-40$A$) will be continued by the
CBM experiment at SIS-300 (FAIR, Darmstadt).
Low cross-section processes, like di-lepton  as well
as open and hidden charm hadron production
will be the focus of these measurements.

Furthermore there is a discussion of the possibility to
construct a heavy ion collider at JINR, Dubna, which
would also cover the low SPS energy range.

\section{Summary}

There are numerous exciting physics questions which motivate
a new experiemntal program to study hadron production in 
hadron-nucleus and nucleus-nucleus collisions at the CERN SPS
which
has been recently proposed by the NA49-future collaboration.
The goals of the program are:
\begin{itemize}
\item
search for the critical point of strongly interacting matter
and a study of the properties of the onset of deconfinemnt
in nucleus-nucleus collisions,
\item
measurements of correlations, fluctuations and hadron spectra at 
high $p_T$ in proton-nucleus collisions needed as for better
understanding of nucleus-nucleus results,
\item
measurements of hadron production in hadron-nucleus interactions
needed for neutrino (T2K) and cosmic-ray (Pierre Auger Observatory and
KASCADE) expriments.
\end{itemize}

The nucelus-nucleus program has the potential for an important
discovery -- the experimental observation of the critical point
of strongly interacting matter. Other proposed studies belong to 
the class of precision measurements.
The collaboartion proposes to perform these measurements in the
period 2007--2011 by use of the upgraded NA49 apparatus.
Synergy of different physics programs as well as the use of the
exisiting accelerator and detectors offer the unique opportunity
to reach the ambitious physics goals in a very efficient and
cost effective way.

\end{document}